\documentclass[sn-basic]{sn-jnl}

\usepackage{graphicx}%
\usepackage{multirow}%
\usepackage{amsmath,amssymb,amsfonts}%
\usepackage{amsthm}%
\usepackage{mathrsfs}%
\usepackage[title]{appendix}%
\usepackage{xcolor}%
\usepackage{textcomp}%
\usepackage{manyfoot}%
\usepackage{booktabs}%
\usepackage{algorithm}%
\usepackage{algorithmicx}%
\usepackage{algpseudocode}%
\usepackage{listings}%

\theoremstyle{thmstyleone}%
\theoremstyle{thmstyletwo}%

\theoremstyle{thmstylethree}%

\begin{document}

\title[]{Implementing Learning Principles with a Personal AI Tutor: A Case Study}


\author[1]{\fnm{Ambroise} \sur{Baillifard}}\email{ambroise.baillifard@unidistance.ch,}
\equalcont{These authors contributed equally to this work.}

\author[2]{\fnm{Maxime} \sur{Gabella}}\email{maxime@magmalearning.com}
\equalcont{These authors contributed equally to this work.}

\author[3]{\fnm{Pamela} \sur{Banta Lavenex}}

\author[3]{\fnm{Corinna S.} \sur{Martarelli}}

\affil[1]{\orgname{EDUDL+, UniDistance Suisse}, \orgaddress{\city{Brig}, \country{Switzerland}}}
\affil[2]{\orgname{MAGMA Learning}, \orgaddress{\city{Lausanne}, \country{Switzerland}}}
\affil[3]{\orgname{Faculty of Psychology, UniDistance Suisse}, \orgaddress{\city{Brig}, \country{Switzerland}}}



\abstract{
Effective learning strategies based on principles like personalization, retrieval practice, and spaced repetition are often challenging to implement due to practical constraints. 
Here we explore the integration of AI tutors to complement learning programs in accordance with learning sciences.
A semester-long study was conducted at UniDistance Suisse, where an AI tutor app was provided to psychology students 
taking a neuroscience course (N=51).
After automatically generating microlearning questions from existing course materials using GPT-3,
the AI tutor developed a dynamic neural-network model of each student's grasp of key concepts. 
This enabled the implementation of distributed retrieval practice, personalized to each student's individual level and abilities. 
The results indicate that students who actively engaged with the AI tutor achieved significantly higher grades. 
Moreover, active engagement led to an average improvement of up to 15 percentile points compared to a parallel course without AI tutor.
Additionally, the grasp strongly correlated with the exam grade,
thus validating the relevance of neural-network predictions.
This research demonstrates the ability of personal AI tutors to model human learning processes and
effectively enhance academic performance. 
By integrating AI tutors into their programs, educators can offer students personalized learning experiences grounded in the principles of learning sciences, thereby addressing the challenges associated with implementing effective learning strategies. 
These findings contribute to the growing body of knowledge on the transformative potential of AI in education.
}




\maketitle

%

\section{Introduction}


Learning is one of the most fundamental human processes. 
One may even say that what makes us human is our extraordinary ability to learn. 
Our mastery of fire, tools, language, etc. is the fruit of our learning, transmitted from generation to generation. 

In recent decades our talent for learning has enabled us to build a technology that is itself able to learn. 
Machine learning, a subset of artificial intelligence (AI), involves the development of computer programs that are capable of improving themselves through experience, or, in other words, learning~\citep{Fleuret2023}. 
Central to this progress is the evolution of artificial neural networks, loosely inspired by the architecture of the human brain. 
These networks consist of interconnected neurons organized into layers, and the learning process involves strengthening specific connections to generate increasingly desirable outputs. 
For example, a classifier network can be trained to identify cats in images. 

The question of how we as human learners should view this newly emerging machine learning
is the subject of much heated debate, often rooted in deep existential fears \citep{munkdebate}.
In this article, however, we adopt a constructive perspective
and consider that the most beneficial application of the irrepressible power of machine learning
is to focus it on understanding how we learn and how we could learn even better. 

The application of AI to education, commonly known as AIEd, has been under active research for nearly three decades (see \cite{Crompton2023} for a recent review). 
However, the success of this approach has fallen short of early expectations. 
Researchers attribute this to the failure of AI-powered learning technologies to take into account solid theoretical foundations established by the learning sciences \citep{Bartolom2018}. 
In a comprehensive review of AIEd, \cite{Zawacki2019} underscored the importance of explicit integration of pedagogical theories in AIEd projects.
Indeed, the field of learning sciences has made invaluable advances in understanding the process of human learning. 
Through empirical and theoretical research, several robust principles have been shown to enhance the effectiveness of learning, including personalization, spaced repetition, and retrieval practice (see for example \cite{Kirschner}).

In this article, we take a step towards bridging the gap between the learning sciences and AIEd by demonstrating how key pedagogical principles can be effectively implemented using machine learning to improve student performance. 
During a one-semester course at UniDistance Suisse, we introduced a personal AI tutor app as a complementary learning activity for students.
The AI tutor first used large language models to generate relevant microlearning questions from course materials. 
Based on gradual interactions with students, a neural network then built a predictive model of their dynamic knowledge levels, in order to adapt the learning process to their individual needs and abilities. 
Our main research question was to assess whether active usage of the AI tutor by students resulted in significantly higher exam grades.
We also investigated the reliability of the neural network's modeling of students' knowledge level, 
as it is a prerequisite for effectively implementing personalized learning.

\section{Reviews}

In this section, we review three concepts that are at the core of the empirical research presented in this article: 
human learning, machine learning, and using machine learning to enhance human learning.

\subsection{Human learning}

The field of learning sciences combines cognitive psychology, neuroscience, and AI to deepen our theoretical understanding of how humans learn and to improve practical learning approaches.
\cite{Dunlosky2013} systematically discussed ten empirically tested learning techniques and evaluated their utility (see also \cite{Weinstein2018}). 
Here we review some of the most effective techniques identified by the learning sciences:
spaced practice, retrieval practice, interleaving, elaboration, and personalization.

\emph{Spaced practice}, also known as \emph{distributed practice}, 
has consistently demonstrated some of the strongest benefits for learning \citep{BENJAMIN2010228}. 
Research has shown that spacing out learning over time benefits long-term retention 
more than does massing learning sessions in close succession.
The robust advantages of spaced practice have been empirically demonstrated across various settings \citep{Toppino2002TheSE, ROEDIGER201120, Jost2021},
and shown to lead to better long-term retention and understanding of learned material.
However, spaced practice requires complex planning and self-awareness, 
as learning material should be reactivated precisely when it has been partially forgotten.
It is indeed believed that letting memory degrade creates a ``desirable difficulty” that helps learning in the long term \citep{bjork2011a}.
Another obstacle when implementing spaced practice is that learning materials are generally arranged in a linear, block-by-block sequence,
which encourages massed learning of each section before moving on to the next one unidirectionally. 

\emph{Retrieval practice} is a well-established learning strategy that involves actively recalling information from memory rather than passively reviewing it
\citep{karpicke2015a, pan2018a}. 
Retrieval practice enhances long-term memory, promotes meaningful learning, and facilitates knowledge transfer to new contexts.
It has proven effective across a wide range of learning situations \citep{ROEDIGER201120}. 
But the benefit of retrieval practice depends on successful retrieval,
as excessively high or low success rates are unlikely to improve memory. 
Moreover, overly difficult retrieval practice exercises may negatively impact students' self-efficacy and confidence \citep{Weinstein2018}.

\emph{Interleaving} enhances learning by alternating between different ideas or concepts, 
as opposed to the common practice of focusing on a single theme during a learning session
(see for example \cite{Rohrer2007}). 
Interleaving promotes broader understanding, discrimination, and the application of efficient strategies.
However, caution is necessary when implementing interleaving, as the relevance and relatedness of the interleaved material can play a significant role \citep{Weinstein2018}.

\emph{Elaboration} involves connecting new information to pre-existing knowledge \citep{Reigeluth1979}. 
It enhances memory retention and understanding by encouraging deeper processing and organization of concepts. 
Elaboration also supports the transfer of knowledge to new situations.
While elaboration offers substantial benefits, implementing it in practice poses challenges such as how to instruct students to elaborate, how to measure the depth of processing, and how to make elaboration time-efficient \citep{Weinstein2018}.

\emph{Personalization} aims to create a flexible learning experience tailored to meet the unique needs of each individual. 
The seminal work of \cite{Bloom1984} reported that students who received personalized one-on-one tutoring performed better than 98\% of students who received uniform training in a group.
Despite high expectations, the realization of personalized tutoring in practice has remained elusive because of its excessive costs. 
In addition, recent experiments and literature on personalization have yielded inconclusive results and revealed pedagogical gaps in implementing personalized approaches \citep{Pane2017, Casta2018}. 

It is worth noting that these techniques can often be advantageously combined. 
In this context, personalization plays a crucial role, 
in particular by determining appropriate spacing intervals for each learner, 
by selecting the desirable levels of difficulty for retrieval practice exercises, 
and by interleaving and connecting concepts in a way that is adapted to individual progress \citep{Bloom1984}. 
Indeed, desirable difficulty depends on individual capacities, preferences, and energy levels, highlighting the significance of personalized learning to enhance the overall learning experience \citep{bjork2011a}.

\subsection{Machine learning}

Artificial intelligence (AI) is a field of research focused on the development and implementation of computer systems capable of performing tasks that typically require human intelligence. 
Recent advances in AI have been largely driven by progress in machine learning (ML), which involves the use of algorithms that can automatically enhance their performance and learn from data. 
By training on large datasets, ML systems can recognize patterns and correlations, enabling them to extract meaningful insights and make accurate predictions. 
This approach has led to significant breakthroughs in various domains such as image and speech recognition, natural language processing, and text generation
(see~\cite{Fleuret2023} for a recent introduction to machine learning).

A fundamental component of machine learning is the concept of artificial neural networks, which draw inspiration from the structure and functioning of the human brain. Neural networks consist of interconnected nodes or ``neurons," usually organized in layers. 
Through a process known as training, these networks strengthen specific connections to produce increasingly desirable outputs. 
This ability to adapt and improve their performance autonomously is a crucial aspect of ML models.

A notable advancement in neural network architecture was the emergence of ``transformer" networks \citep{NIPS2017_3f5ee243}, which have proven to be powerful models for sequence modeling tasks, such as machine translation and document summarization. Transformers excel in natural language processing and understanding by effectively capturing long-range dependencies and contextual information in complex texts. Large language models, which harness the power of transformers and extensive amounts of pre-existing text data, have demonstrated the ability to generate human-like language responses. An example that has garnered significant attention and acclaim is OpenAI's GPT (Generative Pre-trained Transformer) \citep{NEURIPS2020_1457c0d6, openai2023gpt4}.

\subsection{Artificial intelligence in education}

While the application of AI in education (AIEd) has been the subject of research for over three decades, 
recent advancements in machine learning have unlocked a wide range of possibilities to enhance education \citep{Hannele2022, Ouyang2022, Crompton2023}.
These applications can be grouped into four main categories \citep{Zawacki2019}.
(1) Profiling and Prediction: AI algorithms are used to analyze data and create student profiles, enabling timely interventions and predicting outcomes such as admission, academic achievement, and dropout rates.
(2) Assessment and Evaluation: AI-based assessment and evaluation streamline the grading process, offering instant feedback and facilitating comprehensive assessments, including the evaluation of creativity and critical thinking skills.
Instructors can also benefit from AI assistance in generating questions and creating tests.
(3) Adaptive Systems and Personalization: Adaptive systems and personalization  address the limitations of the traditional one-size-fits-all approach. 
These systems provide  learning experiences that cater to individual students' needs and learning abilities, enhancing their effectiveness and engagement.
(4) Intelligent Tutoring Systems: Intelligent tutoring systems leverage AI to provide personalized and adaptive instruction \citep{Elham2021}. 
These systems simulate one-on-one tutoring experiences, tailoring learning materials and feedback to optimize student engagement, knowledge retention, and collaboration. 

The integration of AI in education presents significant pedagogical opportunities by enhancing student support systems, fostering adaptive learning environments, and improving educational practices and outcomes. AI technologies serve as valuable assistants, providing students with personalized support in their zone of proximal development, where they can rapidly develop with appropriate assistance at any time of the day or week~\citep{VYGOTSKY}.


\section{Methods}

\subsection{AI tutor app}

The AI tutor used in this study was a mobile and web application developed by MAGMA Learning \citep{magma}.
It implemented a personalized approach to retrieval practice and spaced repetition,
with the goal of enhancing students' learning effectiveness by consolidating their grasp of key concepts for the long term. 

Before the start of the course in August 2022, a comprehensive set of 800 questions was generated from lecture materials using GPT-3~\citep{NEURIPS2020_1457c0d6} and other natural language processing techniques.
The questions encompassed various formats such as definitions, clozes (fill-in-the-blank), true/false, multiple-choice, image-based, and acronyms. 
The questions also had varying difficulty levels to accommodate the proficiency distribution among students.
Each question was linked to the specific lecture slide that inspired its generation, 
providing students with the relevant contextual feedback if needed. 
After generation, the questions were also individually reviewed and validated by the course instructor. 
Examples of the generated questions can be seen in Figure~\ref{fig:screenshots} as well as in Appendix~\ref{appendixQuestions}.

\begin{figure}[tbh]
\centering
\includegraphics[width=\textwidth]{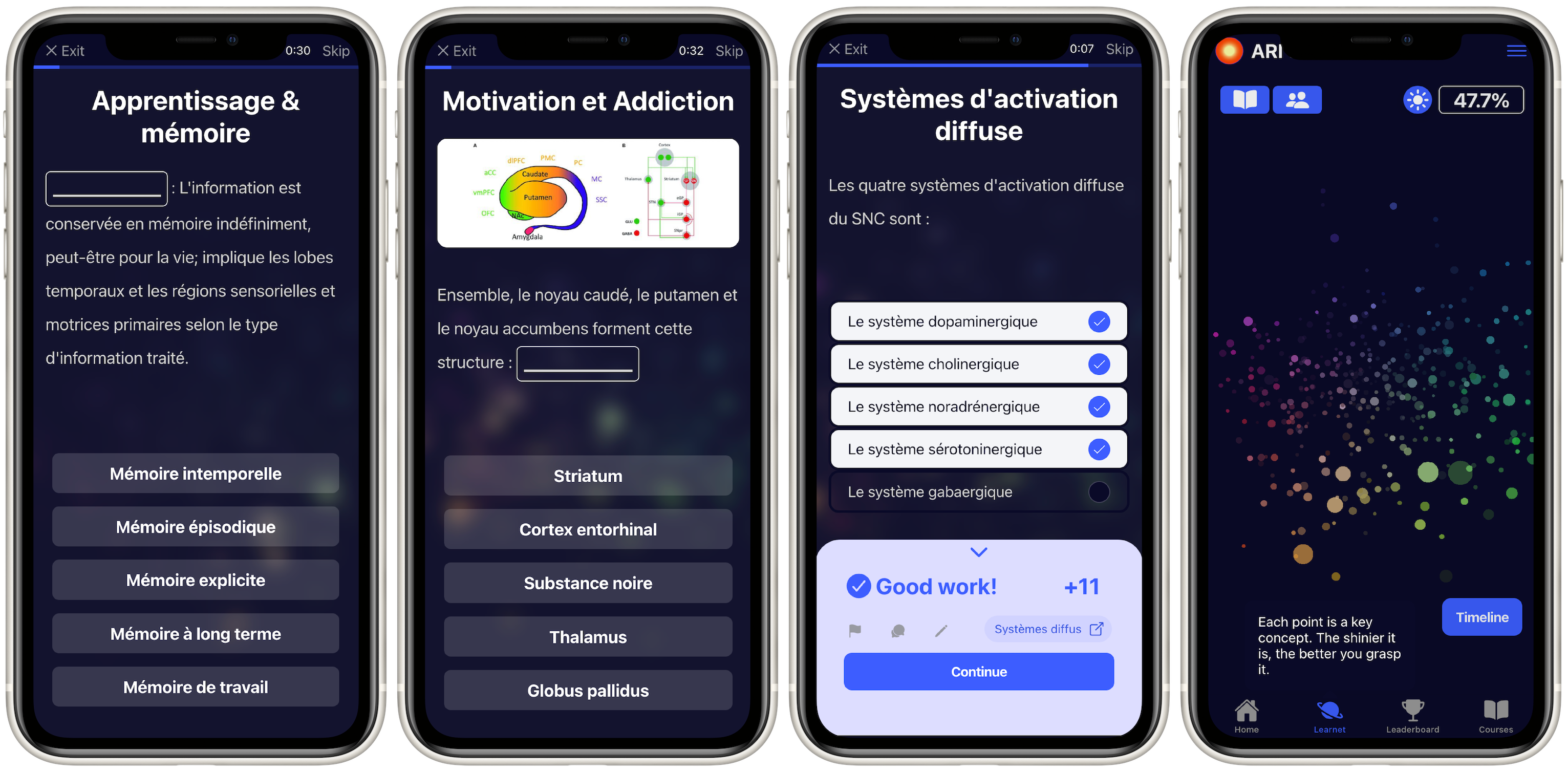}
\caption{Examples of questions generated by the AI tutor app.
From left to right, we see a definition, a question based on an image, and a multiple-choice question with feedback (see also Appendix~\ref{appendixQuestions} for questions in English). On the right we see the ``learnet,'' a visual organization of all key concepts and their grasps by the student (47.7\% in this case). }
\label{fig:screenshots}
\end{figure}

Based on interactions with students and their answers to questions, the AI tutor dynamically 
predicted the probability of a correct answer---referred to as the ``grasp''---for each student and each question.
These predictions were made by an artificial neural network trained with input features derived from information on questions, students, and their historical interactions. 
With this personalized understanding of each individual student's knowledge levels and their evolution (through learning and forgetting), the AI tutor presented the questions considered most relevant and beneficial whenever the student accessed the app. 
The questions that were selected by the app aimed to maintain an appropriate level of challenge for each student, avoiding unstimulatingly easy as well as frustratingly hard questions.
This approach aligns with the concept of ``desirable difficulty" and ensures an engaging learning experience within the student's zone of proximal development~\citep{VYGOTSKY}.

The app provided students with the capability to track their progress through a visual representation of their knowledge called the ``learnet'' (shown on the right of Figure~\ref{fig:screenshots}).
The learnet presented a three-dimensional organization of all the key concepts from the learning materials and their interrelations. 
Each point of the learnet corresponded to a specific course concept,
with its brightness indicating the student's grasp of that particular concept. 
Darker learnets served as motivation for students, indicating areas where they still had more to learn or  had already forgotten. 
Conversely, brighter learnets communicated to students their high knowledge levels.

\subsection{Design}

The study took place during the Fall semester of 2022-2023 in the ``Neuropsychology and Neurosciences” course
offered by UniDistance Suisse as part of the bachelor's curriculum in psychology.
A total of 61 students enrolled in the course, which was taught by one of the co-authors (PBL). 

The course was delivered in a distance learning format,
with online materials and a total of five live and recorded webinars. 
It consisted of 23 lessons grouped into 5 main periods. 
Learning materials were available online on a learning management platform \citep{moodle}. 
Throughout the semester, students had the opportunity to take five quizzes on the Moodle platform, each containing approximately 20 multiple-choice questions, which could be attempted repeatedly for further practice.
The course was worth 10 ECTS credits, corresponding to a workload of 250-300 hours for the students.

On January 28, 2023, a total of 51 students (86\% female;  $\text{mean age}=36.8$, $SD=9.2$) took the final exam for the course. 
The exam consisted of 15 multiple-choice questions. The grading system for the exam ranged from 1 to 6, with a minimum passing grade of 4.
In parallel to the ``Neuropsychology and Neurosciences” course, most students were also enrolled in the parallel ``Neuroanatomy" course, 
and 47 took the exam for the parallel course on the same day. 
Both courses had similar formats and quantities of contents.

The AI tutor app was provided to students as a complementary learning activity for the main  ``Neuropsychology and Neurosciences” course, but not for the parallel ``Neuroanatomy" course. 
The instructor periodically reminded students to use it in order to enhance their acquisition of the learning materials.
Students could use the AI tutor at any time during the semester, but only had access to the questions relevant to the current and preceding periods. 
Students were free to choose the extent to which they used the app, including the frequency, duration, and timing.

\section{Results}

Out of the 51 students who took the final exam for the main course, 43 students could be linked to accounts created on the AI tutor app.
There were 47 students who took both  the ``Neuropsychology and Neurosciences” exam and the parallel ``Neuroanatomy" exam, and among them 40 students were identified as users on the AI tutor app. 
On average, students gave 1800 answers on the app ($SD=2700$) and spent 7.2 hours ($SD=10$) learning on the app, on 26 distinct days ($SD=31$).

We studied the relationship between learning with the personal AI tutor app and 
the grades achieved on the final exam. 
We then explored the causal nature of this relationship by comparing performance
with the parallel ``Neuroanatomy" course (without AI tutor), as well as with another online learning activity (quizzes on Moodle). 
In addition, we investigated the correlation between neural-network predictions of 
students' grasps and their exam grades.

\subsection{Enhanced academic performance}\label{sec:grades}

Among the cohort of 51 students who took the exam for the main course,
43 students signed up on the AI tutor app and used it to varying extents, while 8 students did not sign up. 
Our first objective was to determine whether students who were active on the app got better grades on the final exam than inactive students. 

\begin{figure}[tbh]
\centering
\includegraphics[width=\textwidth]{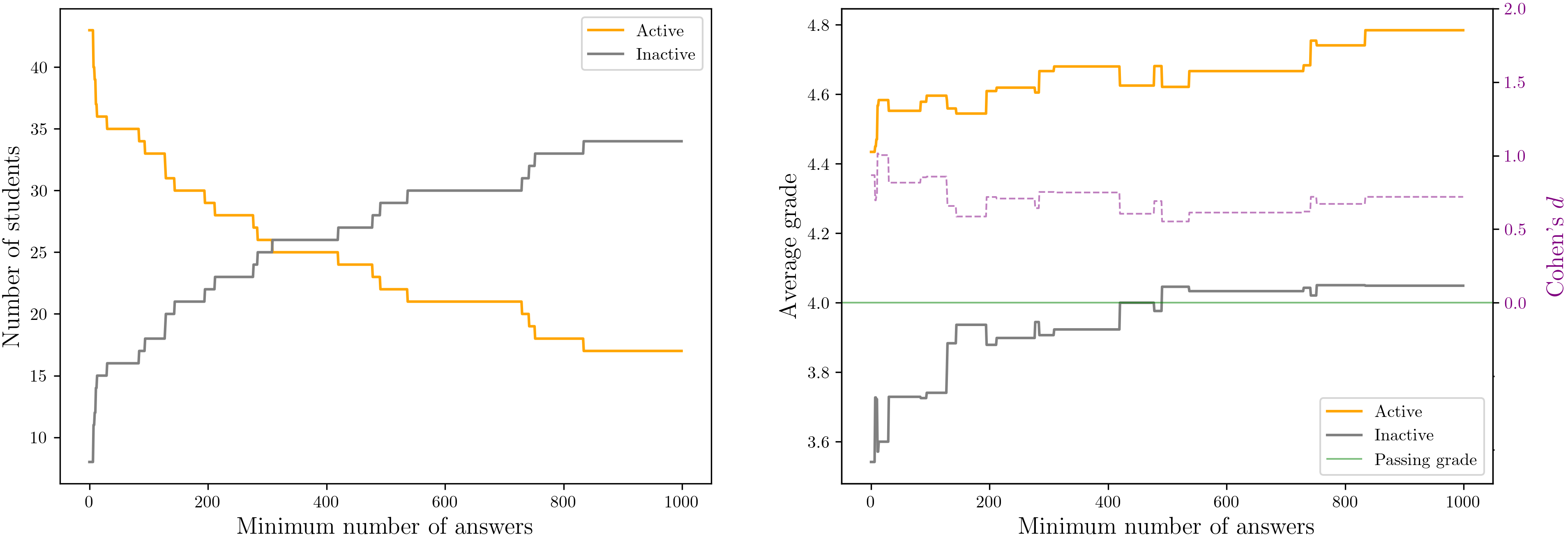}
\caption{\emph{Left-hand side}: Distribution of students ($N=51$) categorized as active or inactive according to the minimum number of answers provided on the app.
\emph{Right-hand side}: Average grades for active and inactive students, as distinguished by a minimum number of answers given on the app. 
The effect sizes (Cohen’s $d$) are indicated on the right vertical axis. 
}
\label{fig:active_grade}
\end{figure}

To define ``active'' participation, we set a minimum threshold for the number of answers provided on the app, denoted as $N_{\text{min}}$.
The inactive group was comprised of students who provided fewer than $N_{\text{min}}$ answers or did not sign up at all.
Conversely, active students were defined as those who provided more than $N_{\text{min}}$ answers.
Although results could a priori depend on the choice of threshold $N_{\text{min}}$, 
we found a significant increase in average grade for the active group,
regardless of the value of the threshold on a large range from $0$ to $1000$, $N_{\text{min}} \in [0,1000]$.
On a scale of 1 to 6, the average grade for the active group was higher by $0.71$ points ($SD=0.08$) compared to the inactive group
(see the evolution of average grades in the right-hand side of Figure~\ref{fig:active_grade}). 

\begin{figure}[tbh]
\centering
\includegraphics[width=0.45\textwidth]{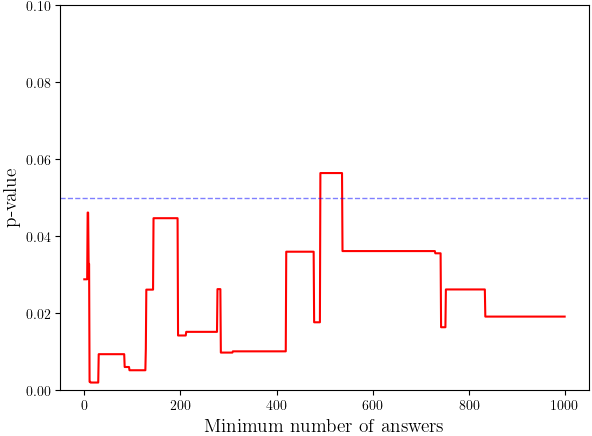}
\caption{$p$-values for the $t$-tests comparing the average grades of active and inactive students
based on the minimum number of answers. The blue dashed line represents the conventional $p$-value threshold of 0.05. }
\label{fig:t_p}
\end{figure}

We found a substantial effect size, reflected by an average Cohen's $d$ of $0.69$ ($SD=0.09$) over the range $N_\text{min} \in [0,1000]$, as illustrated on the right of Figure~\ref{fig:active_grade}.
In order to compare active and inactive students, we conducted independent $t$-tests with 49 degrees of freedom. 
The evolution of the $p$-value with respect to $N_\text{min}$ is depicted in Figure~\ref{fig:t_p}.
Over the range of $N_\text{min} \in [0,1000]$, all $p$-values were found to be below 0.06, with 95\% of them below 0.05. 

\subsection{Comparison to parallel course}\label{sec:ranking}

As previously mentioned, most students enrolled in the main ``Neuropsychology and Neurosciences" course simultaneously took the parallel ``Neuroanatomy" course during the Fall semester of 2022-2023. 
Whereas the AI tutor app was available for the main course, it was not provided for the parallel course. 
This presented an opportunity to investigate whether students who actively learned with the AI tutor for the main course ranked higher (in terms of exam grades) than they did in the parallel course. 

\begin{figure}[tbh]
\centering
\includegraphics[width=0.55\textwidth]{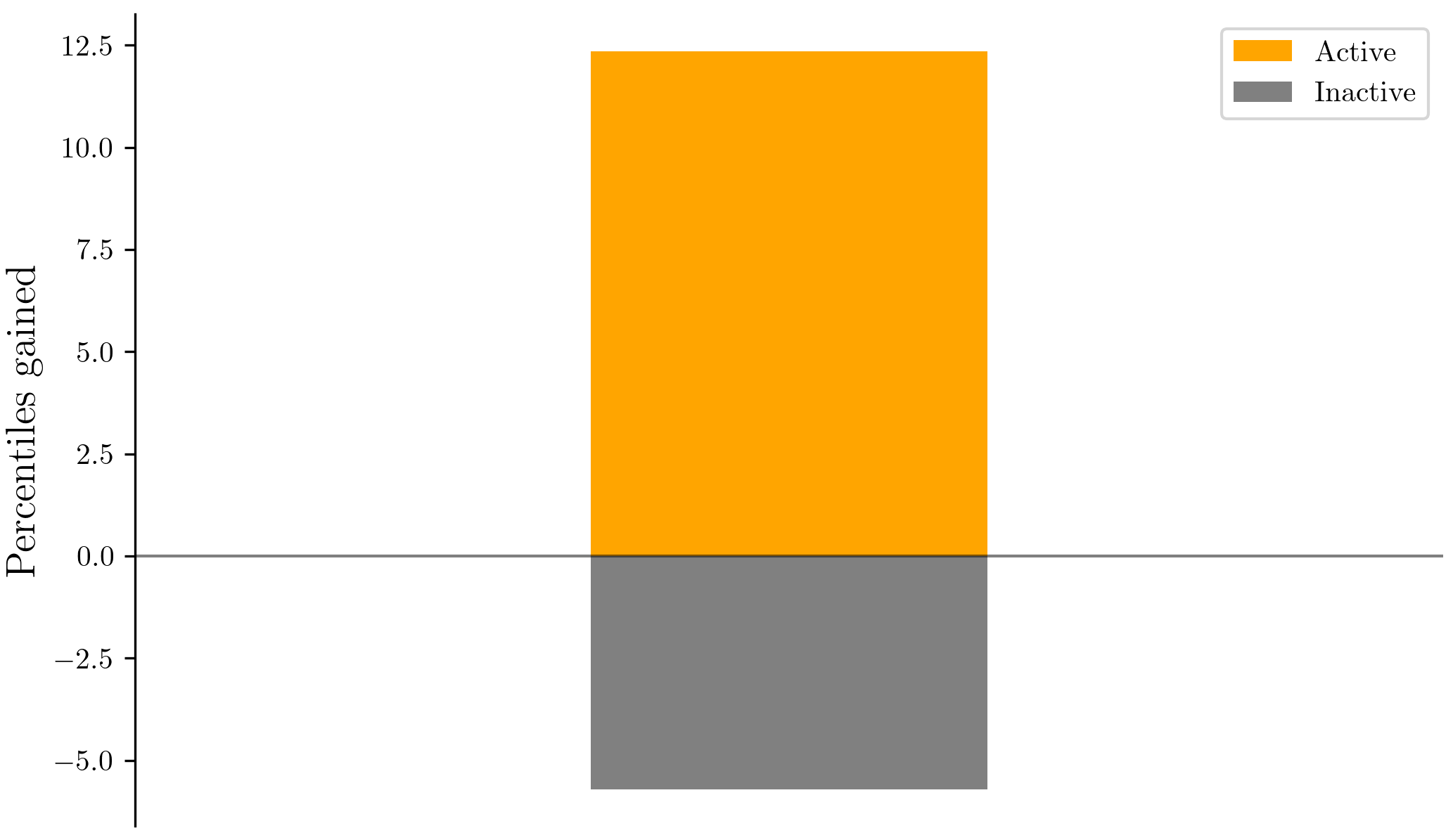}
\caption{
Percentile points gained for the main course, compared to the parallel course ($N=47$). 
Active students (who provided more than 1000 answers on the AI tutor app) gained on average $12.4\%$ percentile points while inactive students lost $5.7\%$. 
}
\label{fig:percentiles}
\end{figure}

For the 47 students who took both the main and the parallel exams, we compared their rankings in terms of percentiles between the two courses.
Active students were defined as those who provided more than $N_{\text{min}}=1000$ answers on the AI tutor app (15 active students).  
Compared to their rankings in the parallel exam, active students gained 18.1 percentile points more than inactive students for the main exam. 
The independent-samples $t$-test gave a $t$-statistic of $t(45) = 2.33$ and a $p$-value of $p=0.025$ 
(Figure~\ref{fig:percentiles}). 
The mean percentile gains were $12.4\%$ for active students and $-5.7\%$ for inactive students, both with a standard deviation of $SD_1 = SD_2 = 25\%$. 
Cohen's $d$ was 0.73. 

\begin{figure}[tbh]
\centering
\includegraphics[width=\textwidth]{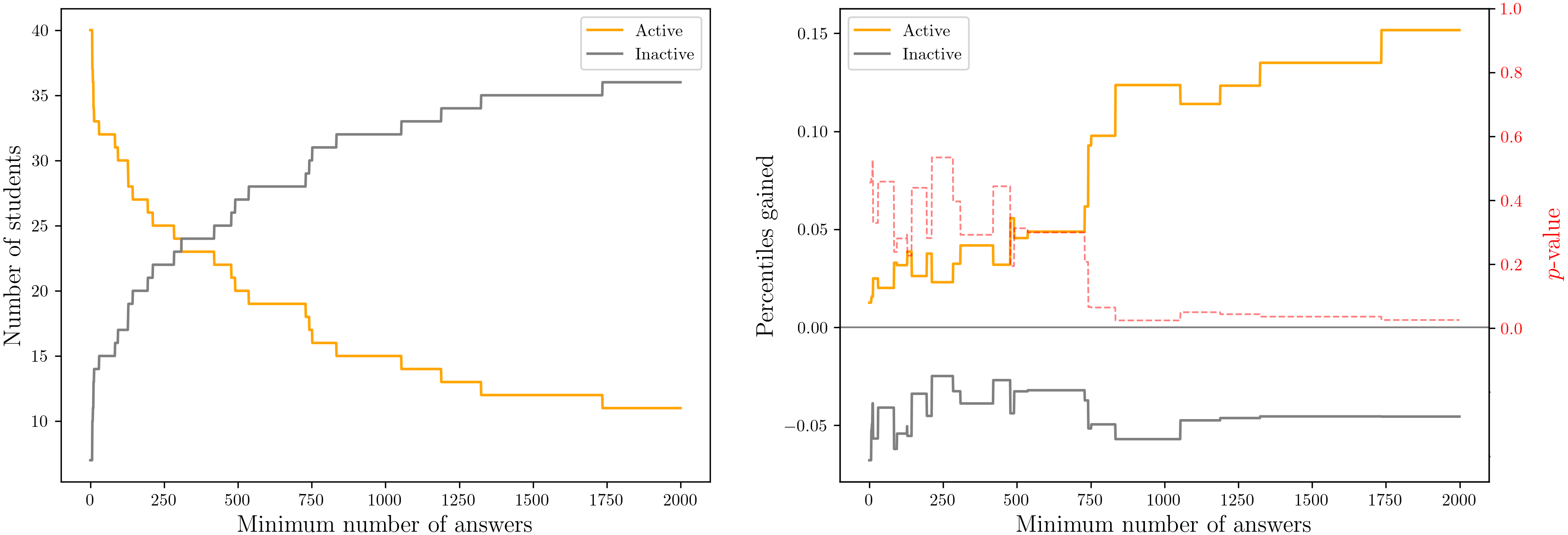}
\caption{
\emph{Left-hand side}: Distribution of students ($N=47$) categorized as active or inactive based on the minimum number of answers provided on the app. 
\emph{Right-hand side}:
Percentile gain from the parallel exam to the main exam. 
Inactive students lost around 5 percentile points regardless of the threshold $N_\text{min} \in [0, 2000]$ while active students gained between 10 and 15 percentile points for $N_\text{min} > 742$.
The $p$-values are indicated on the right vertical axis.}
\label{fig:active_percentile}
\end{figure}

As a next step, we examined the number of students within the active and inactive groups based on the threshold $N_\text{min}$, along with the average percentiles gained for each group.
The results of independent $t$-tests showed a sharp differentiation between the two groups around $N_\text{min} = 750$ answers (Figure~\ref{fig:active_percentile}).
The significance was confirmed by the $p$-values shown on the right of Figure~\ref{fig:active_percentile}. 
While $p$-values for $N_\text{min} < 742$ were all above 0.2, for $N_\text{min} \geq 742$ all $p$-values were below 0.07,
with 82\% below 0.05. 

\subsection{Comparison to Moodle quizzes}\label{sec:moodle}

To further put these results in perspective, recall that students of the main course were provided with 5 quizzes on the Moodle platform. 
Each quiz consisted of around 20 multiple-choice questions covering the different lessons of the course.
Students could take the quizzes multiple times and received feedback with respect to their answers.

\begin{figure}[tbh]
\centering
\includegraphics[width=0.99\textwidth]{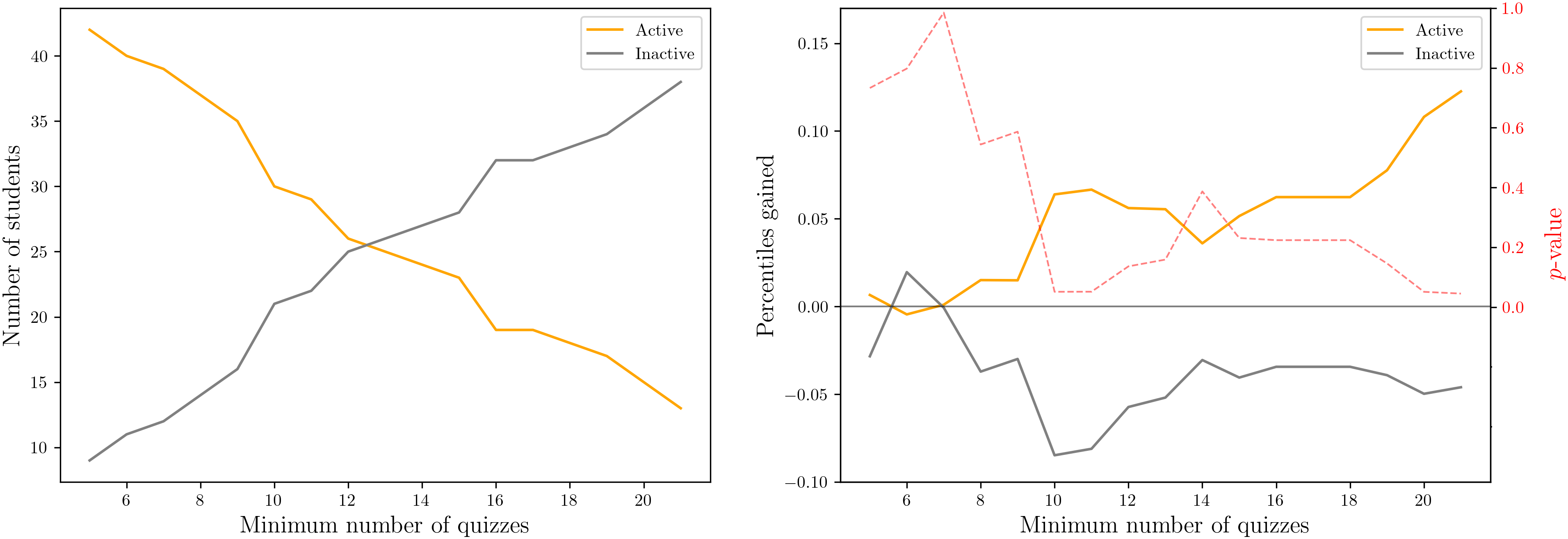}
\caption{
\emph{Left-hand side}: Distribution of students ($N=47$) categorized as active or inactive based on the minimum number of answers provided on the Moodle quizzes. 
\emph{Right-hand side}: Percentile gain from the parallel exam to the main exam. 
Students who took many Moodle quizzes did not get significantly better grades than in the parallel course. Compare with Figure~\ref{fig:active_percentile}.
}
\label{fig:moodle_count_grade}
\end{figure}

We evaluated whether students who engaged in active learning using the Moodle quizzes for the main course achieved higher rankings (in terms of exam grades) compared to their performance in the parallel course. As shown in both descriptive and inferential statistics in Figure~\ref{fig:moodle_count_grade}, taking larger number of quizzes on the Moodle platform did not consistently lead to higher rankings. 
The percentile improvement compared to the parallel course
was not clearly significant, with $p$-values that continued to fluctuate across the range of thresholds.

\subsection{Validation of grasp prediction}\label{sec:grasp}

Finally, we investigated the effectiveness of the AI tutor's neural-network models to meaningfully represent each student's grasp of key course concepts. 
For the AI tutor to effectively personalize the learning experience, it was imperative that these models be realistic.

\begin{figure}[tbh]
\centering
\includegraphics[width=0.45\textwidth]{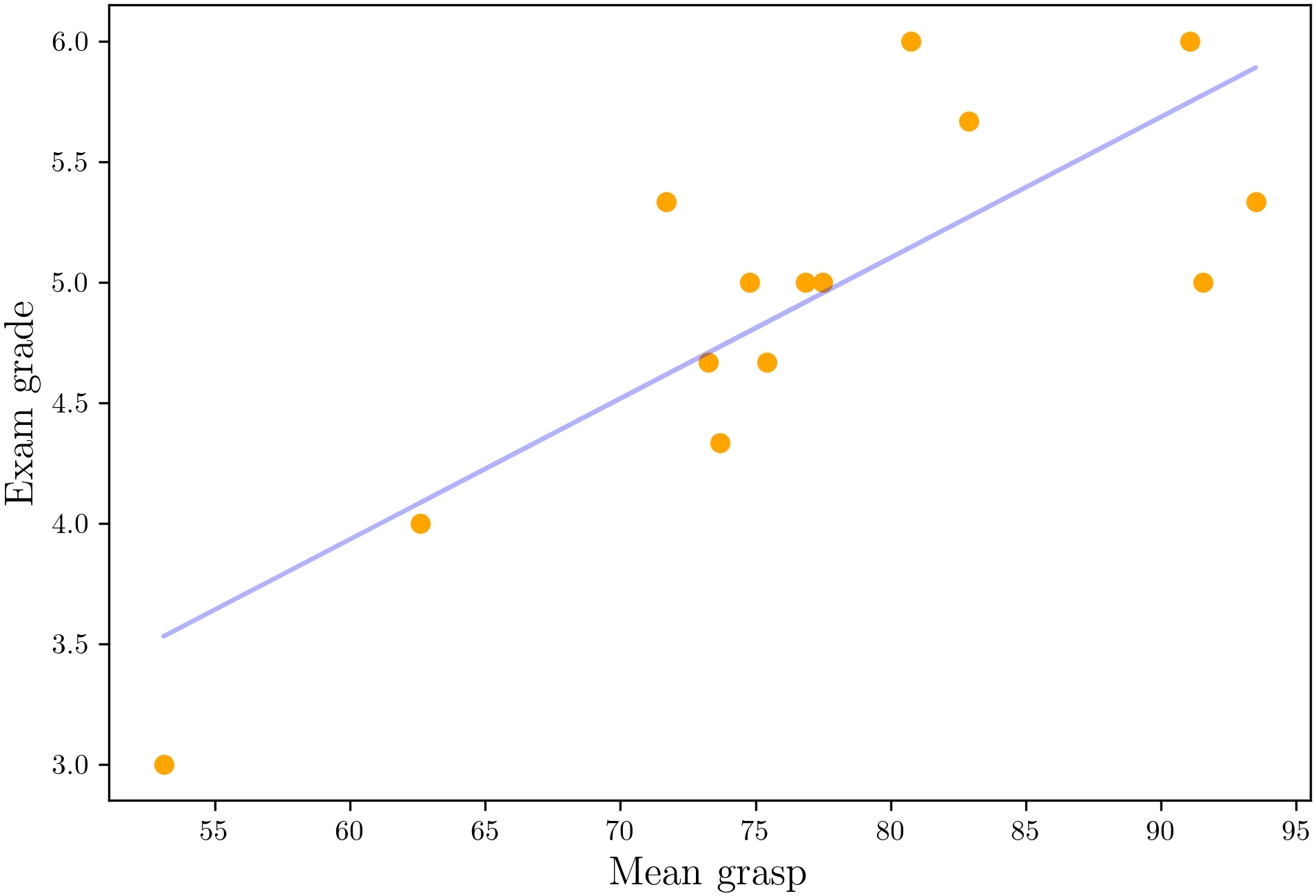}
\caption{Strong correlation ($r = 0.81$) between the integral grasp predicted by the AI tutor and the exam grade, for the group of most regular students ($N=14$).}
\label{fig:mean_grasp}
\end{figure}

To evaluate the relevance of the AI tutor's predictions, we studied the correlation between the predicted grasps (integrated over the duration of the semester) and students' actual grades on the final exam. 
We focused on a group of students who consistently engaged with the app on at least 30 different days throughout the semester (14 students).
Our analysis revealed a strong positive correlation between predicted grasp and exam grade, as evidenced by a high Pearson coefficient of $r(14) = 0.81$, $p< 0.001$, $95\%~\text{Cl}~[0.479, 0.936]$ (Figure~\ref{fig:mean_grasp}).
These results indicated that the AI tutor's neural-network model indeed effectively represented students' knowledge levels. 

\section{Discussion}
 
In this article we evaluated the effectiveness of an artificial intelligence app to implement learning strategies based on
principles validated by the learning sciences. 
We provided psychology students at UniDistance Suisse with a personal AI tutor app for a semester. 

We first found that students who actively engaged with the AI tutor app achieved significantly higher grades on the final exam compared to inactive students (section~\ref{sec:grades}).
This outcome is consistent with previous research underscoring the efficacy of learning principles such as personalization, retrieval practice, and spaced repetition~\citep{Dunlosky2013}.
These principles were effectively incorporated into the design of the AI tutor app.

However, it is important to note that this result in itself does not establish a causal relationship between app usage and improved grades. 
Indeed, it could be that motivated students were more inclined to use the app, and their inherent motivation would have led to superior academic performance in any event, regardless of their app usage~\citep{BUSATO20001057}.
In other words, it may be that app usage and improved grade were both influenced by other variables, 
such as an inherent motivation to study and take part in learning activities. 

Counter-evidence to this possibility was provided by our findings that 
students who learned with the AI tutor app improved their rankings by up to 15 percentile points relatively to the parallel course, for which the AI tutor app was not provided (section~\ref{sec:ranking}).
If active students obtained better grades only because they were more motivated (both to use the app and to pass the exam), then their motivation would have likely resulted in superior performance in the parallel course as well, and therefore their relative improvements would average to zero.
Note that further evidence for the positive impact of learning with the AI tutor 
comes from the fact that significant effects on percentile gains only start to appear above a threshold of around 750 answers provided by students. 
This is precisely what one could expect to observe if the effect was due to the AI tutor app, since the number of available questions was of the same order (800). 

Still, an explanation for why students only improved in the main course (with the AI tutor) could be that 
these students simply were more engaged in that course than in the parallel course (without AI tutor). 
If so, their engagement would have been a common cause for their 
high levels of participation in all learning activities (including using the AI tutor app)
as well as for their superior grades in the main course. 
However, if this were the case, then we would expect that another learning activity, namely Moodle quizzes, 
should also lead to consistently significant percentile gains, which was not observed (section~\ref{sec:moodle}). 

In sum, all of our findings converge to suggest a beneficial impact of 
learning with the personal AI tutor app on academic performance. 
We attribute this success to the capacity of the app's neural networks to model human learning processes in a
meaningful way, as suggested by the strong correlation between the predicted grasp and the exam grade
 (section~\ref{sec:grasp}).
 
As with all intervention studies, appropriate control groups and conditions are difficult to design and interpret when studying the effectiveness of learning apps. 
One limitation of our study is that we did not incorporate an active control group with random assignment of participants. 
We chose not to adopt this approach in our design, as the group of students without access to the app might have felt disadvantaged.
Indeed, academic institutions have the responsibility to provide all of their students with the same learning activities and opportunities.
Nevertheless, our statistical analyses allowed us to go beyond simply observing the relationship between app usage and exam grade. 
We were able to provide additional corroborating evidence by comparing the performances of the same group of students across different courses and online learning activities. 

One interesting direction for future research would consist in providing a personal AI tutor app to a group of students, 
but only activating personalized learning functionalities for half of them. 
A preliminary unpublished work \citep{Alemanno2022}
showed that the group of students with deactivated personalization
performed increasingly poorly on the app and eventually disengaged completely. 
Research on a larger scale could help to confirm the findings presented here and identify precisely what learning behaviors with the AI tutor app enhance performance the most. 

We have presented strong arguments in favor of the effectiveness of 
implementing learning principles using a personal AI tutor, using natural language processing and machine learning. 
In contrast to most studies on the subject, the study presented here was not reductionist, 
but rather corresponded to realistic learning conditions that lasted an entire semester. 
Given the wide range of learning activities in any single course (lectures, reading, quizzes, etc.), it is surprising and promising that the addition of an AI tutor can have such significant and beneficial effects.

%
%
%
%
%
%
%
%
%
%

\section*{Declarations}

\begin{itemize}
\item Funding

This research was funded by UniDistance Suisse.  

\item Conflict of interest

One of the co-authors (MG) is the CEO of MAGMA Learning, the company that developed the AI tutor app studied in this article. 

\item Ethics approval 

Data usage was approved by UniDistance Suisse. Since the data are archival and anonymous, no written informed consent was required. 

\end{itemize}

\begin{appendices}

%
%
%
%
%

\section{Examples of microlearning questions}\label{appendixQuestions}

{\parindent0pt 

{\bf Question: \,}
The \underline{\hspace{1.5cm}} envelops \underline{\hspace{1.5cm}} in \underline{\hspace{1.5cm}} and stamps them with a tag indicating the location where they are to be transported.

{\bf Correct answers: \,}
Golgi apparatus, proteins, vesicles.

{\bf Distractors: \,}
enzymes,
ribosomes,
lipids,
cytoskeletal elements.

\

{\bf Question: \,}
The four diffuse modulatory systems of the central nervous system are: 

{\bf Correct answers: \,}
Cholinergic system,
Serotonergic system,
Noradrenergic system,
Dopaminergic system.

{\bf Distractors: \,}
Glutamatergic system,
Gabaergic system.

\

{\bf Question: \,}
The hormone produced by the neurohypophysis that is implicated in reproduction, coupling, orgasm, parentality, etc. 

{\bf Correct answers: \,}
Oxytocin.

{\bf Distractors: \,}
Prolactin,
Luteinizing hormone,
Follicle-stimulating hormone,
Oxycontin.

\

{\bf Question: \,}
The model of associative \underline{\hspace{1.5cm}} that is based on long-term \underline{\hspace{1.5cm}} describes the \underline{\hspace{1.5cm}} of an \underline{\hspace{1.5cm}} between distinct stimuli that are perceived simultaneously. 

{\bf Correct answers: \,}
learning,
potentiation,
formation,
association.

\

{\bf Question: \,}
The inhibition of inhibitory \underline{\hspace{1.5cm}} interneurons in the \underline{\hspace{1.5cm}} leads to an \underline{\hspace{1.5cm}} in the liberation of dopamine in the \underline{\hspace{1.5cm}} of the ventral striatum.

{\bf Correct answers:  \,}
GABAergic,
ventral tegmental area,
increase,
nucleus accumbens.

{\bf Distractors:  \,}
glycinergic,
decrease,
amygdala.

\

{\bf Question (Figure~\ref{fig:screenshots}): \,}
\underline{\hspace{1.5cm}}: Information is kept in memory indefinitely, perhaps for life; involves the temporal lobes and primary sensory and motor regions depending on the type of information processed. 

{\bf Correct answers: \,}
Long-term memory.

{\bf Distractors: \,}
Timeless memory,
Episodic memory,
Explicit memory,
Working memory.

\

{\bf Question (Figure~\ref{fig:screenshots}): \,}
Together, the caudate nucleus, the putamen and the nucleus accumbens form this structure: \underline{\hspace{1.5cm}}

{\bf Correct answers: \,}
Striatum.

{\bf Distractors: \,}
Entorhinal cortex,
Substantia nigra,
Thalamus,
Globus pallidus.

\

{\bf Question (Figure~\ref{fig:screenshots}): \,}
The four diffuse activation systems of the CNS are:

{\bf Correct answers: \,}
The dopaminergic system,
The cholinergic system,
The noradrenergic system,
The serotonergic system.

{\bf Distractors: \,}
The gabaergic system.

}

\end{appendices}


\bibliography{sn-bibliography}

\end{document}